\documentclass[iop]{emulateapj}
\usepackage{color}

\newcommand{\NuSTAR}{\textit{NuSTAR}}
\newcommand{\Swift}{\textit{Swift}}
\usepackage{ulem}
\usepackage{amsmath}
\usepackage{enumerate}
\usepackage[colorlinks=true,citecolor=blue,breaklinks=true]{hyperref}
\newcommand\cts{counts~s$^{-1}$}

\begin{document}

\title{The spin of the black hole GRS~1716-249 determined from the hard intermediate state}

\author{Lian Tao\altaffilmark{1}, John A. Tomsick\altaffilmark{2}, Jinlu Qu\altaffilmark{1}, Shu Zhang\altaffilmark{1}, Shuangnan Zhang\altaffilmark{1,3,4}, Qingcui Bu\altaffilmark{1}}

\altaffiltext{1}{Key Laboratory of Particle Astrophysics, Institute of High Energy Physics, Chinese Academy of Sciences, Beijing 100049, China}
\altaffiltext{2}{Space Sciences Laboratory, 7 Gauss Way, University of California, Berkeley, CA 94720-7450, USA}
\altaffiltext{3}{National Astronomical Observatories, Chinese Academy of Sciences, Beijing 100012, China}
\altaffiltext{4}{University of the Chinese Academy of Sciences, Beijing, China}

\shorttitle{GRS~1716-249}
\shortauthors{Tao et al.}

\begin{abstract}

We present three simultaneous/quasi-simultaneous \textit{NuSTAR} and \textit{Swift} datasets of the black hole GRS~1716-249 in its hard intermediate state. The accretion disk in this state may have reached the innermost stable circular orbit, and the \textit{NuSTAR} spectra show a broad relativistic iron line and a strong Compton hump. To measure the black hole spin, we construct a joint model consisting of a relativistic disk model {\tt kerrbb} and a reflection model {\tt relxill}, to fit the continuum and the reflection components, respectively. By applying this model to each dataset independently, a consistent result is obtained on the black hole spin and the disk inclination. The black hole spin is $a^{\ast} \gtrsim 0.92$, and the inclination angle ($i$) is around 40--50$^{\circ}$, based on the measurements of all datasets. In the third dataset, a high black hole mass ($M_{\rm BH}$) is strongly disfavored by the spectral fits. By unfreezing the black hole mass, we find $a^{\ast}>0.92$, $i=49.9^{+1.0\,\circ}_{-1.3}$ and $M_{\rm BH}<8.0$\,$M_\odot$, at a 90\% confidence level. Considering the lower limit derived from a previous optical constraint, $M_{\rm BH}$ is in a range of $4.9-8.0$\,$M_\odot$.

\end{abstract}

\keywords{accretion, accretion disks --- black hole physics --- individual (GRS~1716-249) --- X-rays: binaries}

\section{Introduction}
\label{sec:intro}

A black hole (BH) can be described by its spin and mass. Many physical processes, such as the formation of black holes, the growth of seed black holes via accretion, and the launching of powerful jets, are closely related to spin \citep[for a review see][]{Middleton2016}.

BH spin can be measured by looking for the signatures of spin on X-ray spectra. The reflection \citep[e.g.,][]{Fabian1989,Reynolds2014} modeling and continuum \citep{Zhang1997,McClintock2014} modeling are two techniques developed to measure BH spin. The key step in both techniques is to measure the location of the inner disk radius. Assuming that the accretion disk extends to the innermost stable circular orbit (ISCO), we can infer the spin. A smaller ISCO indicates a faster BH spin: for a maximally prograde rotating BH, the ISCO is at 1\,$R_{\rm g}$, where $R_{\rm g}$ is the gravitational radius; for a non-spinning BH, the ISCO expands to 6\,$R_{\rm g}$; for a maximally retrograde rotating BH, the ISCO is at 9\,$R_{\rm g}$.

In the continuum modeling, there are three basic assumptions. First, the inner disk radius should be at the ISCO; second, the accretion disk is a standard thin disk, which means that the disk luminosity varies as the inner disk temperature raised to the fourth power; finally, the BH mass, distance and disk inclination angle are already known. The orbital inclination angle is usually used instead of the disk inclination angle. However, the orbital planes of some sources may be misaligned with their inner disk planes, such as Cygnus X-1 \citep{Tomsick2014,Walton2016}, GRO J1655-40 \citep{Hjellming1995} and SS 433 \citep[e.g.,][]{Blundell2004}. So using the orbital inclination angle can result in an incorrect BH spin measurement.

If the BH spin is measured through the reflection modeling, high quality spectra are required. The spectra should have broad energy coverage, including the iron line ($\sim 6$~keV) and the Compton hump ($\sim 20-40$~keV). In addition, pileup effect can distort the iron line profile and lead to an incorrect spin value \citep[e.g.,][]{Miller2010}, so the spectra should be pileup-free. Data from \textit{Nuclear Spectroscopic Telescope Array} \citep[\textit{NuSTAR;}][]{Harrison2013} are sensitive to the reflection component and are free from pileup effect, thus they are particularly suited for the reflection modeling. Moreover, as a bonus, the disk inclination angle can be obtained in the spectral fitting.

If one source can reach a special state, in which the standard accretion disk extends to the ISCO and the reflection and continuum components are also significant, the spin will be well constrained using both the continuum and reflection modelings simultaneously. In this case, the inclination is not needed to be known in advance, as it can be obtained from the reflection modeling. This method has been successfully used in the BH spin measurement of Cygnus X-1 \citep{Tomsick2014}.

Black hole GRS~1716-249 has reached such a state during its 2016--2017 outburst \citep{Bassi2019}. When the source was in the hard intermediate state, the disk luminosity varied with the fourth power of the inner disk temperature, consistent with the scenario of the standard thin disk. In addition, the inner disk radius ($R_{\rm in}$) in this state was constant, suggesting that the accretion disk may extend to the ISCO. Moreover, the emission contribution from the corona was significant, so a strong disk reflection component is also expected in the spectrum. In this context, GRS~1716-249 is a good target to perform the spin measurement, following the method above.

Here we also summarize other basic properties of GRS~1716-249. It is an X-ray transient discovered with the \textit{Compton Gamma Ray Observatory} (\textit{CGRO})/BATSE \citep{Harmon1993} and \textit{Granat}/SIGMA \citep{Ballet1993} in the 1993 outburst. The source is a low mass X-ray binary (LMXB), with a main sequence companion star of spectral type $K$ or later \citep{della Valle1994,Masetti1996}. The distance is estimated to be $2.4 \pm 0.4$\,kpc \citep{della Valle1994}. The BH mass, derived from the superhump period of 14.7~hr, is larger than $4.9\,M_\odot$ \citep{Masetti1996}.

In this paper, we report on simultaneous/quasi-simultaneous \textit{NuSTAR} and \textit{Swift} observations taken in the hard intermediate state of the 2016--2017 outburst. The BH spin is measured via the joint modeling of the continuum and the reflection components. This paper is organized as follows. A description about observations and data reduction are presented in Section~\ref{sec:obs}, the detailed spectral fittings are presented in Section~\ref{sec:res} and the results are discussed in Section~\ref{sec:dis}.

\begin{deluxetable*}{llclll}
\tablecolumns{6}
\tablewidth{\textwidth}
\tabletypesize{\scriptsize}
\tablecaption{\NuSTAR\ and \Swift\ Observations \label{tab:obs}}
\tablehead{
\colhead{Mission} & \colhead{ObsID}  & \colhead{Start Time} & \colhead{End Time} & \colhead{Exposure} & \colhead{Count rate} \\
      &    &    &    &  \colhead{(s)} & \colhead{(\cts)}
}
\startdata
\multicolumn{6}{c}{Dataset 1}\\
\NuSTAR & 90202055002 & 2017-04-07~14:26:09 & 2017-04-08~03:11:09   & 17897.5 & $152.68 \pm 0.09$ \\
\Swift  & 00034924029 & 2017-04-07~08:50:41 & 2017-04-07~09:18:56   & 1687.0 & $110.7\pm 0.4$ \\
        & 00034924030 & 2017-04-08~23:11:36 & 2017-04-08~23:27:56   & 978.1 & $103.5\pm 0.5$ \\
\noalign{\smallskip}\hline\noalign{\smallskip}
\multicolumn{6}{c}{Dataset 2}\\
\NuSTAR & 90202055004 & 2017-04-10~16:36:09 & 2017-04-11~03:36:09 & 15797.5 & $149.04 \pm 0.10$ \\
\Swift  & 00034924031 & 2017-04-10~21:12:42 & 2017-04-10~23:05:56 & 1945.0 & $95.5\pm 0.3$ \\
\noalign{\smallskip}\hline\noalign{\smallskip}
\multicolumn{6}{c}{Dataset 3}\\
\NuSTAR & 90301007002 & 2017-07-28~12:06:09 & 2017-07-30~23:21:09 & 89256.0 & $52.63 \pm 0.02$ \\
\Swift  & 00034924051 & 2017-07-29~20:00:17 & 2017-07-29~20:16:55 & 993.1 & $79.2\pm 0.3$ \\
        & 00034924052 & 2017-07-30~18:30:24 & 2017-07-30~18:46:56 & 988.0  & $74.4\pm 0.3$ 
\enddata
\tablecomments{For \NuSTAR\ observations, the exposure and the net count rates of FPMA are listed. For \Swift\ observations, the net count rates in grade 0 have been corrected for hot columns, bad pixels and loss of counts caused by pileup, using the PSF correction factor in {\tt xrtmkarf}. }
\end{deluxetable*}

\begin{figure}
\centering
\includegraphics[width=\columnwidth]{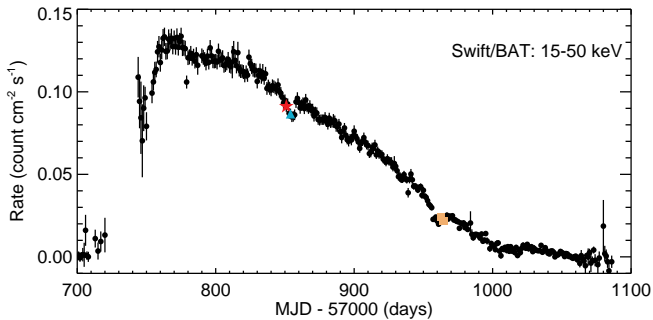} \\
\caption{X-ray light curve of GRS~1716-249 observed with \textit{Swift}/BAT in the 15--50~keV band. The overlaps of the three \textit{NuSTAR} observations are indicated by red star, blue triangle and yellow squares, respectively.   
\label{fig:obs}}
\end{figure}

\section{Observations and Data reduction}
\label{sec:obs}

Based on the spectral state classification in \citet{Bassi2019}, three \textit{NuSTAR} observations are taken in the hard intermediate state, with an exposure from 16~ks to 89~ks (see Figure~\ref{fig:obs} and Table~\ref{tab:obs}). Simultaneous \textit{Swift} observations are also collected\footnote{For the \textit{NuSTAR} observation of ObsID 90301007002, there are three simultaneous \textit{Swift} observations (ObsID 00088233001, 00034924051 and 00034924052). The spectrum of ObsID 00088233001 is different from the two others, we thus discard this observation. Moreover, we also test the joint fits below (Section~\ref{sec:res}) including this observation, but the results just change marginally.}, in order to extend energies below $1$~keV, except one \textit{NuSTAR} observation (ObsID 90202055002). For this \textit{NuSTAR} observation, there are no strict simultaneous \textit{Swift} observations. We thus use two quasi-simultaneous observations instead, whose time differences are less than 20~hr.

The cleaned \textit{NuSTAR} event files are processed using NuSTARDAS pipeline 1.8.0, with the \textit{NuSTAR} CALDB version 20180419. The count rate of two observations (ObsID 90202055002 and 90202055004) exceed 100~\cts, then some source events may be discarded improperly in {\tt nupipeline}. Therefore, for the two observations, we create their cleaned event files by setting the keyword {\tt statusexpr} to be ``STATUS==b0000xxx00xxxx000'', following the \textit{NuSTAR} analysis guide\footnote{https://heasarc.gsfc.nasa.gov/docs/nustar/analysis/}. The spectra are then extracted by the {\tt nuproducts} script from a circular region. The region centers at the source and has a radius of 180\arcsec. Another nearby source-free circular region is used as the background region. The spectra are rebinned to have at least 50 counts per energy bin using {\tt grppha}.

All \textit{Swift} observations are taken in windowed timing mode. The cleaned event files are created by {\tt xrtpipeline} with XRT CALDB version 20180710. Using XSELECT in HEASoft v6.24, the source and background spectra are extracted from grade 0 events, respectively. The source extraction region is a circular region with a radius of 20 pixels, if the observations are not affected by pileup. The background extraction region centered at the source position, with the inner radius of 90 pixels and the outer radius of 110 pixels. Pileup are present in four observations, whose count rate are larger than 90~\cts\ (see Table~\ref{tab:obs}). Therefore, for these observations, following \citet{Bassi2019}, the central regions with a radius of 3 pixels are excluded when extracting source spectra. Finally, the spectra are rebinned to contain minimum 50 counts per bin.

\begin{figure}
\includegraphics[width=0.45\textwidth]{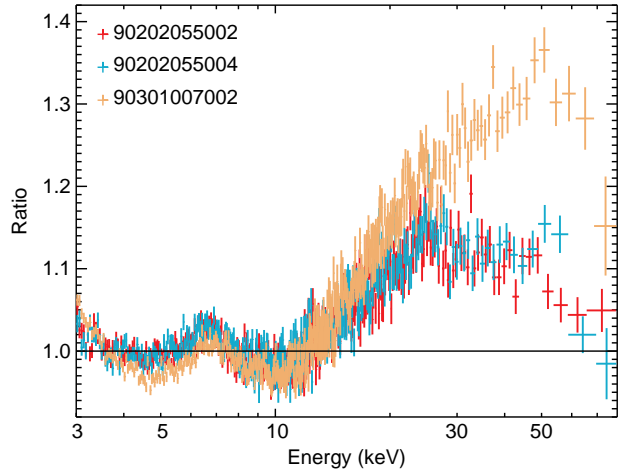} 
\caption{Data to model residuals for the three \textit{NuSTAR} observations when fitting the spectra in 3-4, 9-11 and 40-79 keV bands with an absorbed cutoff powerlaw model. ObsID 90202055002, 90202055004 and 90301007002 are marked in red, blue and yellow, respectively. The plots are grouped to have a signal-to-noise ratio (S/N) of at least 50 per bin for display clarity, and only the FPMA data are shown here. 
\label{fig:iron}}
\end{figure}

\section{Spectral modeling and Results}
\label{sec:res}

XSPEC v12.10.0e software package \citep{Arnaud1996} is used to do the spectral fits. As shown in Figure~\ref{fig:iron}, a clear iron line and a strong Compton hump are present in each \textit{NuSTAR} observation, when fitting the spectra in the energy intervals of 3--4, 9--11, 40--79 keV with an absorbed cutoff powerlaw model. The reflection component is significant, and the accretion disk may extends to the ISCO, as suggested by the \textit{Swift} monitoring observations \citep{Bassi2019}\footnote{\textit{Swift} can monitor the whole outburst and observe the source at soft X-rays ($< 1$\,keV), it can thus provide a better understanding for the state transition and the spectral properties of the accretion disk. So, we use the state classification and the accretion disk parameter ($R_{\rm in}$) from \citet{Bassi2019} in this paper.}. Therefore, we can perform the joint modeling of the reflection and the continuum components to measure the BH spin.

\begin{figure*}
\includegraphics[width=0.99\textwidth]{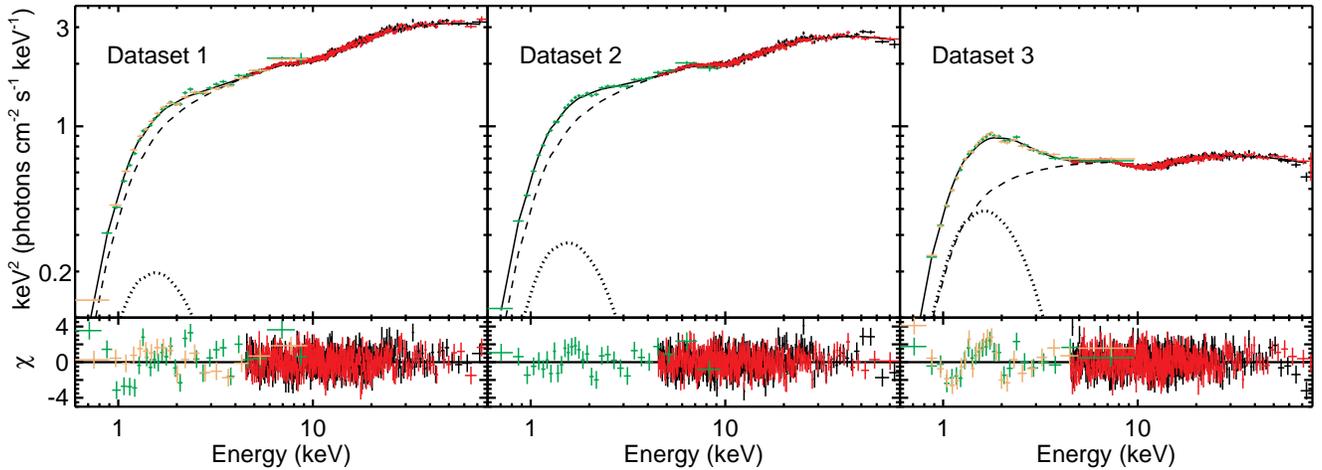} 
\caption{Top: energy spectra and model components when $M_{\rm BH}$ is fixed at 5\,$M_\odot$. The total model, the {\tt relxill} model and the {\tt kerrbb} model are marked in black solid lines, dashed lines and dotted lines, respectively. Bottoms: spectral residuals with respect to the best-fit model. The plots are grouped to have a S/N$\ge 50$ per bin for display clarity. 
\label{fig:fit1}}
\end{figure*}

We use v1.2.0 of the reflection model {\tt relxill} \citep{Garcia2014,Dauser2014} to fit the reflection component, and use the relativistic accretion disk model {\tt kerrbb} \citep{Li2005} to fit the continuum component simultaneously. The interstellar absorption is simulated by {\tt tbabs}, with the abundances of \citet{Wilms2000} and the cross-sections of \citet{Verner1996}. In the {\tt kerrbb} model, we fix the distance at 2.4\,kpc, and link the inclination angle and the spin to that of {\tt relxill}. The spectral hardening factor ($f_{\rm col}$) is fixed at the default value, 1.7\footnote{$f_{\rm col}=1.7$ is the recommended value for a low accretion rate \citep{Shimura1995} and is the mean value for accretion disk around a stellar-mass black hole (https://heasarc.nasa.gov/xanadu/xspec/manual/XSmodelKerrbb.html). So, we use $f_{\rm col}=1.7$ in our case.}. The BH mass is larger than $4.9\,M_\odot$ \citep{Masetti1996}, thus the fits are tested with a BH mass of $5\,M_\odot$, $10\,M_\odot$, $15\,M_\odot$, respectively. In {\tt relxill}\footnote{A high-energy cutoff powerlaw ({\tt cutoffpl}) continuum is the intrinsic incident spectrum in {\tt relxill}, see http://www.sternwarte.uni-erlangen.de/$\sim$dauser/research/relxill/.}, the inner disk radius ($R_{\rm in}$) is fixed at $-1$, which means that the accretion disk extends to the ISCO, as suggested by \citet{Bassi2019}. The inner and outer emissivity indices are fixed at the default values ($q_{\rm in}=3$, $q_{\rm out}=3$). If the emissivity indices are allowed to vary freely, we find that the indices are negative in some fits, which means that the emissivity increases with radius and thus is physically unreasonable. \textit{Swift} data in the 0.6--10 keV band and \textit{NuSTAR} data in the 4.5--79.0 keV band are used. \textit{NuSTAR} data below 4.5 keV are discarded to avoid the mismatch between \textit{Swift} and \textit{NuSTAR} at low energies. The mismatch has also been reported in several other sources by the \textit{NuSTAR} instrument team and may be due to the presence of a dust scattering halo in these sources\footnote{See http://iachec.scripts.mit.edu/meetings/2019/presentations\\/WGI\_Madsen.pdf.}. By comparison, the mismatch in GRS~1716-249 could also possibly be a result of a dust scattering halo. Good fits are obtained for all three datasets, and the results are shown in Figure~\ref{fig:fit1} and Table~\ref{tab:fit1}. In all fits, the BH spin is high ($a^{\ast} \gtrsim 0.92$), and the inclination angle is around 40--50$^{\circ}$.


Recently, \citet{Bharali2019} approximated the same data in dataset 1\footnote{For dataset 1, \citet{Bharali2019} used the \textit{Swift} observation of ObsID 00034924029, but we use one more \textit{Swift} observation, ObsID 00034924030.} and 2, and fitted the spectra with a lamp-post reflection model ({\tt relxilllpCp}) in the {\tt relxill} model family and a {\tt diskbb} model. By fixing $a^{\ast} = 0.998$, the authors claimed that the accretion disk was truncated. However, as we know, the {\tt diskbb} model is a simple model which neglects relativity, and is inappropriate to make physical conclusions based on the {\tt diskbb} parameters when the powerlaw component dominates the spectra. In the datasets presented here, the powerlaw component contributes most of the flux (see Figure~\ref{fig:fit1}), so the {\tt diskbb} model may not be an ideal choice. But in order to compare with \citet{Bharali2019}, we still adopt a {\tt diskbb} model in below tests. Using the same data with same energy coverage\footnote{\citet{Bharali2019} used an energy coverage of 0.5--8.0\,keV for \textit{Swift} and 3.0--79.0\,keV for \textit{NuSTAR}. The fits here are performed with the same energy coverage for consistency. We also test the fits with 0.6--10.0\,keV for \textit{Swift} and 4.5--79.0\,keV for \textit{NuSTAR}. With a better fit (e.g., $\chi^2/{\rm dof}=3075.2/2951$ versus 2892.5/2872 for dataset 2), the main results will not change, and $R_{\rm in}$ of the {\tt relxill} model are even smaller ($1.6^{+0.7}_{-0.6}$, $1.4^{+1.3}_{-0.4}$ and $1.5^{+0.3}_{-0.5}$\,$R_{\rm ISCO}$ for dataset 1, 2 and 3, respectively), consistent with an untruncated or a slightly truncated disk.} as \citet{Bharali2019}, we carefully perform spectral fits with the {\tt relxilllpCp} + {\tt diskbb} model or with the {\tt relxill} + {\tt diskbb} model, and test two cases, i.e., by fixing $a^{\ast} = 0.998$ and allowing $R_{\rm in}$ to vary or by fixing $R_{\rm in} = -1$ and allowing $a^{\ast}$ to vary. Dataset 3, which has the longest \textit{NuSTAR} observation ($\sim 89$\,ks) and is not reported in \citet{Bharali2019}, is also used.

In above trials when fixing $R_{\rm in}$ at the ISCO and allowing $a^{\ast}$ to vary, we note that the spin can not be well constrained if the {\tt diskbb} model is used, regardless of the reflection model ({\tt relxill} or {\tt relxilllpCp}). If we use the {\tt kerrbb} model to fit the disk component, the spin can be well constrained. In comparison to the {\tt diskbb} model, the {\tt kerrbb} model is a more physical model taking full relativistic effects (e.g., BH spin) into account, thus has power to constrain BH spin.

Also, we find that the reflection model used ({\tt relxilllpCp} or {\tt relxill}) will affect the measurement of the inner disk radius, and both {\tt relxilllpCp} and {\tt relxill} have a similar goodness-of-fit, $\chi^2/{\rm dof} = 3302.8/3122$ versus 3292.1/3121 for dataset 1 and $\chi^2/{\rm dof} = 3071.9/2952$ versus 3075.2/2951 for dataset 2. The disk is just slightly truncated with $R_{\rm in} = 6.8^{+1.9}_{-1.8}$ and $2.8^{+2.7}_{-0.8}$\,$R_{\rm ISCO}$ for dataset 1 and 2 when the {\tt relxill} model is used. For dataset 3, where we do clearly see the thermal component, $R_{\rm in}$ (=$1.3^{+0.5}_{-0.3}$\,$R_{\rm ISCO}$) is consistent with the disk being at the ISCO or being slightly truncated if using the {\tt relxill} model, and the {\tt relxill} model provides a much better fit than that of the {\tt relxilllpCp} model ($\chi^2/{\rm dof} = 3952.2/3543$ versus 4183.6/3544). Moreover, in the case of dataset 1, even for the {\tt relxilllpCp} model, the assumption that the disk is at the ISCO also can not be excluded in \citet{Bharali2019}. In their Figure 7, $\Delta \chi^2$ is less than 0.5 for $R_{\rm in} = 1$\,$R_{\rm ISCO}$. In addition, the spectral fits of a variable $R_{\rm in}$ ($a^{\ast} = 0.998$) is not better than that of a fixed $R_{\rm in}$ (a variable $a^{\ast}$) for the {\tt relxill} model. $\chi^2/{\rm dof}$ of dataset 1, 2 and 3, respectively, are 3292.1/3121, 3075.2/2951 and 3952.2/3543 for the former case, but have similar values of 3292.6/3121, 3075.4/2951 and 3952.4/3543 for the latter case. From above tests, we conclude that an untruncated disk is still a reasonable assumption. Especially for dataset 3, which has the strongest thermal component, the results are compatible with a disk at the ISCO, allowing for a constraint on the BH spin.

\begin{deluxetable*}{llll|lll|lll}
\tablecolumns{10}
\tablewidth{0pc}
\tablewidth{\textwidth}
\tablecaption{Spectral parameters of the hard intermediate state \label{tab:fit1}}
\tablehead{
\colhead{Para.} & \multicolumn{3}{c}{Dataset 1} & \multicolumn{3}{c}{Dataset 2} & \multicolumn{3}{c}{Dataset 3} 
}
\startdata
\multicolumn{10}{c}{{\tt tbabs}} \\
$N_{\rm H}$ & $0.690^{+0.011}_{-0.010}$  &  $0.738^{+0.012}_{-0.010}$ & $0.760\pm 0.012$ &$0.67\pm 0.02$ & $0.713^{+0.015}_{-0.013}$& $0.744^{+0.016}_{-0.014}$ & $0.738^{+0.011}_{-0.012}$ & $0.779^{+0.010}_{-0.009}$& $0.849\pm 0.009$\\
\noalign{\smallskip}\hline\noalign{\smallskip}
\multicolumn{10}{c}{{\tt kerrbb}} \\ 
$M_{\rm BH}$ &  5  &  10 & 15 & 5 & 10 & 15 & 5 & 10 & 15   \\
$\dot{M}$  & $3.2^{+0.7}_{-0.3} $  & $2.8^{+1.2}_{-0.3}$ & $3.7^{+0.7}_{-0.4}$  &$5.3^{+1.8}_{-1.4}$ & $5.2^{+1.0}_{-0.5}$ & $6.3^{+0.9}_{-0.6}$ & $9.3\pm0.7$ & $6.9\pm 0.2$ & $8.3\pm 0.2$   \\
\noalign{\smallskip}\hline\noalign{\smallskip}
\multicolumn{10}{c}{{\tt relxill}}\\
$a^{\ast}$ & $>0.988$  &  $>0.981$ & $>0.992$  &$>0.93$ & $>0.991$& $>0.994$ & $0.933^{+0.013}_{-0.011}$ & $>0.9977$& $>0.9978$  \\
$i$  & $43^{+3}_{-2}$  & $ 52.9^{+1.2}_{-1.7}$ & $53.2^{+1.2}_{-1.3}$  &$43^{+3}_{-7}$ & $46^{+2}_{-3}$& $48 \pm 2$&$49.4^{+1.1}_{-1.2}$ & $51.3^{+0.8}_{-0.9}$& $51.2 \pm 0.6$  \\
$\Gamma$  & $1.661^{+0.010}_{-0.014}$  & $1.773^{+0.007}_{-0.017}$ & $1.772^{+0.008}_{-0.010}$  &$1.71^{+0.02}_{-0.05}$ & $1.735^{+0.012}_{-0.018}$ & $1.747^{+0.011}_{-0.015}$ &$1.933^{+0.007}_{-0.016}$ & $1.927^{+0.006}_{-0.004}$ & $1.921\pm 0.002$ \\
${\rm log}(\xi)$ & $3.45^{+0.09}_{-0.07}$  & $3.04^{+0.05}_{-0.03}$ & $3.04^{+0.04}_{-0.03}$  &$3.53^{+0.44}_{-0.11}$ & $3.46^{+0.13}_{-0.08}$ & $3.43^{+0.09}_{-0.07}$ & $3.45^{+0.10}_{-0.05}$ & $3.54^{+0.07}_{-0.06}$& $3.691^{+0.017}_{-0.074}$ \\
$A_{\rm Fe}$  & $3.4^{+0.8}_{-0.6}$  & $1.01^{+0.29}_{-0.05}$ & $1.00^{+0.18}_{-0.06}$  &$4.3^{+5.7}_{-1.3}$ & $3.0^{+1.0}_{-0.6}$ & $2.5^{+0.5}_{-0.4}$ & $3.3^{+0.8}_{-0.3}$ & $3.5^{+0.3}_{-0.4}$ & $3.69^{+0.14}_{-0.11}$ \\
$E_{\rm cut}$  & $300^{+70}_{-50}$  & $1000^{+0}_{-170}$ & $1000^{+0}_{-90}$  &$250^{+70}_{-50}$ & $360\pm 80$& $430\pm 70$ & $1000^{+0}_{-300}$ & $1000^{+0}_{-30}$& $1000^{+0}_{-14}$ \\
$R_{\rm ref}$  & $0.125^{+0.021}_{-0.019}$  & $0.146^{+0.010}_{-0.006}$ & $0.148^{+0.011}_{-0.007}$  &$0.105^{+0.027}_{-0.018}$ & $0.14^{+0.03}_{-0.02}$ & $0.16\pm 0.02$ & $0.135^{+0.013}_{-0.017}$ & $0.167^{+0.016}_{-0.015}$& $0.192^{+0.018}_{-0.019}$ \\
$N_{\rm rel}$  & $2.63\pm 0.09$  & $3.20^{+0.06}_{-0.10}$ & $3.18^{+0.07}_{-0.09}$  &$2.41\pm 0.08$ & $2.39\pm 0.10$& $2.39\pm 0.11$ & $0.85\pm 0.03$ & $0.80\pm 0.03$& $0.77\pm 0.03$ \\
\noalign{\smallskip}\hline\noalign{\smallskip}
$\chi^2/{\rm dof}$ & 3587.8/3462 &   3606.1/3462 & 3616.6/3462 & 2894.9/2873 & 2905.0/2873 & 2922.4/2873 & 3727.6/3460 & 3790.5/3460 & 4050.1/3460
\enddata
\tablecomments{\\
$N_{\rm H}$ is the X-ray absorption column density in units of $10^{22}~\rm atoms~cm^{-2}$; \\
$M_{\rm BH}$ is the BH mass in units of $M_\odot$; \\
$\dot{M}$ is the effective mass accretion rate of the disk in units of $10^{15}~\rm g~s^{-1}$; \\
$a^{\ast}$ is the BH spin in dimensionless units; \\
$i$ is the disk inclination angle in units of deg; \\
$\Gamma$ is the power-law photon index of the incident spectrum; \\
$\xi$ is the ionization parameter of the accretion disk in units of $\rm erg~cm~s^{-1}$;\\
$A_{\rm Fe}$ is the iron abundance in units of solar abundance; \\
$E_{\rm cut}$ is the observed cutoff energy of the incident spectrum; \\
$R_{\rm ref}$ is the reflection fraction; \\
$N_{\rm rel}$ is the normalization of {\tt relxill} model in an order of $10^{-2}$; \\
All errors are quoted at 90\% confidence level.}
\end{deluxetable*}

\section{Discussion}
\label{sec:dis}
We present the simultaneous/quasi-simultaneous \textit{NuSTAR} and \textit{Swift} observations of GRS~1716-249 in the hard intermediate state of 2016-2017 outburst. In this state, the accretion disk extends to the ISCO and the reflection components are significant in all three datasets. Therefore, we can perform the joint modeling of the continuum and the reflection components to measure the BH spin. By modeling the spectra from 0.6--79\,keV with a relativistic disk model {\tt kerrbb} and a reflection model {\tt relxill} simultaneously, and linking the spin and the inclination of {\tt kerrbb} to that of {\tt relxill}, we have measured the BH spin and confirmed the source to be a high spin BH ($a^{\ast} > 0.92$). 

The spectra exhibit some variations in different datasets, consistent with the spectral evolution reported by \citet{Bassi2019}. Following their \textit{Swift} spectral and timing analysis results, the third dataset is taken in the softest episode. Indeed, the incident spectrum of the third dataset is softest ($\Gamma \sim 1.93$) and the disk component is strongest (Figure~\ref{fig:fit1}). The observed cutoff energy ($E_{\rm cut}$) can not be well constrained due to the limited energy coverage of \textit{NuSTAR} ($\lesssim 79$\,keV), so we will not discuss the evolution of $E_{\rm cut}$ in this paper.

Although some variations between the different datasets are observed, three key parameters, i.e., BH spin, disk inclination angle and iron abundance, show good consistency in all datasets, suggesting that our results are reliable. We also note that there are some exceptions. In the first dataset, if freezing $M_{\rm BH}$ at 10 or 15\,$M_\odot$, $A_{\rm Fe}$ is less than other fits. Actually, there are two local minima in the fitting processes which favor a high $A_{\rm Fe}$, and the increase in $\Delta\chi^2$ is 7.6 and 24.1 for $M_{\rm BH}=$ 10 and 15\,$M_\odot$, respectively. In the two local best-fits, the BH spin approaches the maximum value (i.e., 0.998), and the inclination angle is around $47-49^{\circ}$. Therefore, the change in $A_{\rm Fe}$ does not significantly affect the determination of the spin and the inclination angle.

The BH spin is found to be larger than 0.92, and the disk inclination angle $i$ is about 40--50$^{\circ}$, and the iron abundance is $A_{\rm Fe}\sim 3-4$ (in solar units), considering the results of all three datasets. The inclination measurement agrees with the fact that no dips and/or eclipses are observed in both \textit{Swift}/XRT and \textit{INTEGRAL}/JEM-X light curves \citep{Bassi2019}. LMXBs are expected to show dips and/or eclipses in the light curves if the inclination angle is larger than 60$^{\circ}$ \citep{Frank1987}. The spin measurement is also consistent with the result of \citet{Bassi2019}. Assuming $M_{\rm BH}>4.9$\,$M_\odot$ and $i < 60^{\circ}$, they inferred $a^{\ast}>0.8$ from the inner disk radius by modeling the disk component with a simple disk model {\tt diskbb}. 

\begin{figure*}
\includegraphics[width=0.99\textwidth]{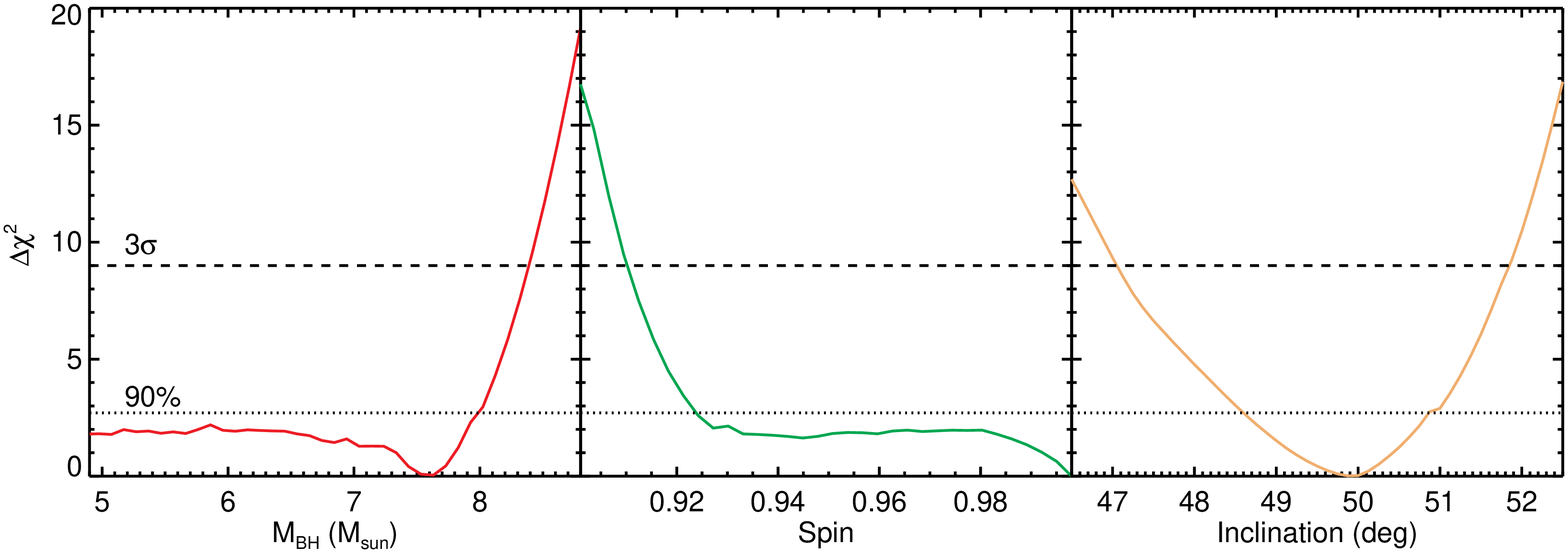} 
\caption{The $\Delta \chi^{2}$ confidence contours of the BH mass (left), the dimensionless BH spin (middle) and the disk inclination angle (right), respectively. The dashed and dotted lines indicate 3$\sigma$ and 90\% confidence levels, respectively. 
\label{fig:cont1}}
\end{figure*}

Only the lower limit of the BH mass is known ($M_{\rm BH}>4.9$\,$M_\odot$), therefore the spectral fitting are tested with $M_{\rm BH}=5$, 10 and 15\,$M_\odot$, respectively. In all three datasets, $M_{\rm BH}$ of 5\,$M_\odot$ provides an overall best-fit. By setting $M_{\rm BH}=10$ or 15\,$M_\odot$, $\chi^2$ will increase at least 10 (see Table~\ref{tab:fit1}). Especially in the third dataset, $\chi^2$ significantly increase as the $M_{\rm BH}$ increases. We thus test the spectra fitting by thawing $M_{\rm BH}$ in the third dataset. $M_{\rm BH}$ in the best-fit model is 7.6\,$M_\odot$, with a reduced $\chi^2$ of 3725.8 for 3459 degrees of freedom. At a 90\% confidence level (Figure~\ref{fig:cont1}), $M_{\rm BH}$ is $7.6^{+0.4}_{-2.7}$\,$M_\odot$\footnote{We set the lower limit of $M_{\rm BH}$ to be $4.9$\,$M_\odot$.}, $a^{\ast}>0.92$ and $i=49.9^{+1.0\,\circ}_{-1.3}$. At a 3$\sigma$ confidence level, the BH mass is $M_{\rm BH}=7.6^{+0.8}_{-2.7}~M_\odot$, and $a^{\ast}$ is larger than 0.91, and the inclination angle is $i=49.9^{+1.9\,\circ}_{-2.8}$. We also test the fits by fixing $M_{\rm BH}$ at $7.6$\,$M_\odot$ for dataset 1 and 2. At a 90\% confidence level, $a^{\ast}$ are $>0.993$ and $> 0.987$, and $i$ are $46^{+2\,\circ}_{-3}$ and $45^{+2\,\circ}_{-3}$, for dataset 1 and 2, respectively, similar to the results of dataset 3.

In summary, performing joint fits of the continuum and reflection components, we measure some basic parameters (spin, inclination and mass) of the black hole GRS~1716-249. The results indicate that the source is a fast spinning black hole with $a^{\ast} \gtrsim 0.92$, $i=$40--50$^{\circ}$ and $M_{\rm BH} = 4.9-8.0$\,$M_\odot$.

\acknowledgements

We thank the anonymous referee for useful comments that have improved the paper. We thank Javier A. Garc{\'{\i}}a for his help in using the {\tt relxill} model. LT acknowledges funding support from the National Natural Science Foundation of China (NSFC) under grant number U1838115, the CAS Pioneer Hundred Talent Program Y8291130K2 and the Scientific and technological innovation project of IHEP Y7515570U1. SZ thanks support from the NSFC under the grant No. U1838201. 

{\it Facilities:} \NuSTAR, \Swift

\end{document}